\documentclass[11pt]{article}
\usepackage{graphicx}
\usepackage{setspace}
\usepackage{empheq}
\usepackage{cite}
\usepackage{amsmath}
\usepackage{dutchcal}
\usepackage{textcomp}
\usepackage{amssymb}
\usepackage{authblk}

\usepackage{color}
\definecolor{MyLinkColor}{rgb}{0,0,0.4}

\newcommand{\ov}{\overline}

\newcommand{\p}{\partial}

\newcommand{\e}{\varepsilon}

\newcommand{\sign}{\mathop{\rm sign}\nolimits}

\newcommand{\f}{\frac}

\numberwithin{equation}{section}

\begin{document}
\title{On the time-evolution of resonant triads in rotational capillary-gravity water waves}


\author[$\dagger$]{Rossen I. Ivanov}
\author[$\star$]{Calin I. Martin}

\affil[$\dagger$]{School of Mathematical Sciences, Technological University Dublin, City Campus, Kevin Street, Dublin D08, NF82, Ireland}
\affil[$\star$]{Faculty of Mathematics, University of Vienna, Oskar-Morgenstern-Platz 1, 1090 Vienna, Austria}

\maketitle

 \begin{abstract}
 We investigate an effect of the resonant interaction in the case of
 one-directional propagation of capillary-gravity surface waves arising as the free surface of a rotational water flow.
 Specifically, we assume a constant vorticity in the body of the fluid which physically corresponds to an underlying current with a linear horizontal velocity profile.
 We consider the interaction of three distinct modes and we obtain the dynamic equations for a resonant triad. Setting the constant vorticity equal to zero we recover the well known
  integrable three-wave system.
\end{abstract}

\bigbreak
\bigbreak


\section{Introduction}

The mathematical study of water waves, or, more generally, of fluids with free surface is an active research area since the 19th century. We aim in this work at one important aspect of the water wave propagation, namely \emph{resonant interactions} of surface waves. Indeed, resonant interactions greatly impact the evolution of waves through the
significant energy transfer among the dominant wave trains, un upshot of the latter process being the appearance of
wave patterns that are higher and steeper than those covered by the linear wave theory. On the other hand, resonant interactions are effective in explaining the outburst of weak turbulence \cite{PusZak}. For valuable insights into the onset
and evolution of turbulence in shear flows we refer the reader to the works by Eckhardt \cite{Eck1999, Eck2011, Eck2018}.

Past studies \cite{Phil,MG} pertaining to irrotational water flows have shown that three-wave resonant interactions are not possible for purely gravitational water waves, but could take place for the capillary-gravity waves when the surface tension is taken into account and the initial waves are not collinear. For the one-directional wave propagation, we would like to note that, while capillary waves in irrotational flow do not exhibit three-wave resonances \cite{KAMS}, it was shown in \cite{CK} that special constant vorticities trigger the appearance of resonances. Moreover,  the combined action of capillarity and gravity also gives rise to three-wave resonances in the case of one-dimensional propagation of rotational water waves of constant vorticity \cite{MarJmfm}.

However, besides being a generator of resonances, the vorticity is also a cause for major complications which significantly diminish the analytical tractability. Indeed, the presence of swirling motions  impedes the utilization
of velocity potentials very much used in studies dealing with irrotational flows. Nevertheless, rigorous mathematical studies
\cite{CoSt2004, ConsSaSt, CoEsch} persistently improved the understanding of water waves with vorticity which enframes
inter-related fundamental oceanic phenomena (like upwelling/downwelling, zonal depth-dependent currents with flow reversal). The importance of vorticity in the realistic modeling of ocean flows is underlined in recent studies
by Constantin and Johnson \cite{CJPhysFluids2017, CJOcean, CJJPhysOc19} and Constantin and Monismith \cite{CoSm}.

Experimental studies on resonant interactions are reported in \cite{MG2}. For the one-dimensional propagation this system is integrable by the well known inverse scattering method \cite{ZMNP,ZM}.

In this study we derive the three-wave nonlinear system for the one-directional propagation of capillary-gravity  water waves in the case of a rotational fluid.
More specifically, we assume a constant vorticity in the body of the fluid. It is to note that flows with constant vorticity are representative of a wide range of physical settings.
Indeed, constant vorticity accommodates an underlying current with a linear horizontal velocity profile. Moreover, if the waves are long compared with the average water
depth, then the existence of non-zero mean vorticity is more important than its specific distribution, cf. the discussion by Teles da Silva and Peregrine \cite{daS}, so that constant vorticity becomes relevant.
A constant vorticity distribution also describes tidal currents, with positive/negative vorticity being appropriate for the ebb and flood, respectively.
The importance of these aspects is underlined by the fact that in areas of the continental shelf and in many coastal inlets these are the most significant currents
(see the discussion in the paper \cite{CSV} by Constantin, Strauss and Varvaruca).

In such a constant vorticity setting we consider the interaction of three distinct modes and derive the dynamical equations for a resonant triad.
We would like to emphasize that, when setting the constant vorticity equal to zero, our equations recover the nonlinear system describing the three-wave interaction in the irrotational case from \cite{CaCh}.

Concerning the four-wave interaction, we would like to note that evolution characteristics of an integrable system of four
waves above irrotational flow were performed by Stiassnie and Shemer \cite{Stia} by employing the Zakharov equation
\cite{Zak} and the deep-water dispersion relation for gravity waves. A statistical analysis of the evolution of four-wave interaction based on the Zakharov equation is presented in \cite{StSt}.
 For experimental and numerical investigations concerning four-wave interactions over underlying currents we refer the reader to \cite{Was}.

We note that there are studies on the interaction of an infinite number of modes, leading to the concepts of energy cascades and wave turbulence, (see e.g. \cite{Busta}, the review paper \cite{Kart} and the references therein) however we do not explore this direction. For a comprehensive presentation of analytical, computational and more applied aspects concerning water wave resonances we refer the reader to the works of Bustamante and Kartashova \cite{BuKa09, BuKa09sec, BuKa11, Busta}.

 In our derivation we are using the Hamiltonian approach for the capillary-gravity waves. Following the seminal paper of Zakharov \cite{Zak}, the Hamiltonian approach for water waves propagation has been developed extensively, see for example \cite{Broer,Mil1,Mil2, BO,Rad}. A convenient explicit representation of the Hamiltonian involves the non-local Dirichlet-Neumann operator,  e.g. \cite{CraigGroves1,CGK}. In some particular cases rotational fluids, such as fluids with constant (or piecewise) vorticity can also be treated within the Hamiltonian framework \cite{NearlyHamiltonian, Wahlen, ConsIvPoF, CIM, ConsIvCMP}.

\section{The physics of the problem}
\subsection{The governing equations}
\begin{figure}[h!]
\centering
\includegraphics[width=0.8 \textwidth]{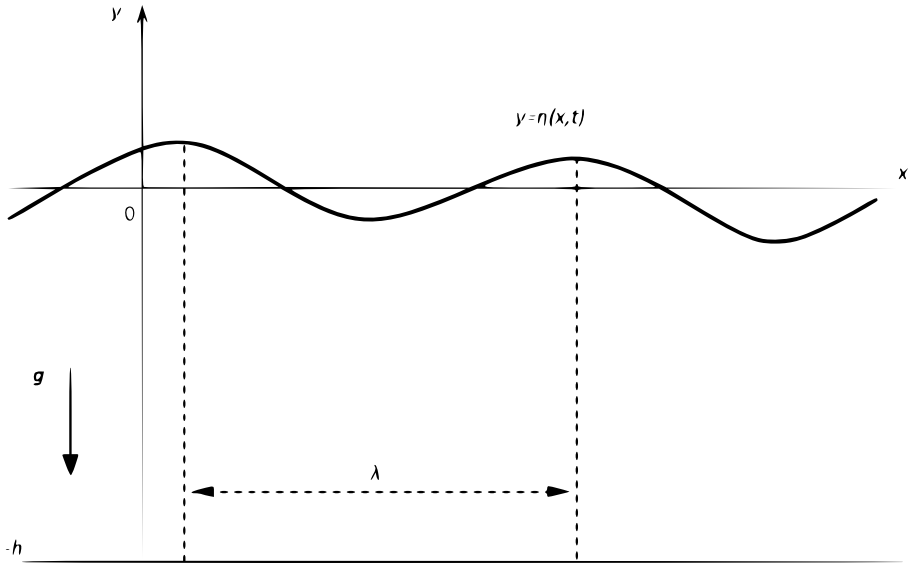}
 \caption{The setup of the wave propagation.}\label{fig1}
\end{figure}

First we choose a $xy$ coordinate system, where $x$ denotes the horizontal variable and $y$ stands for the vertical
variable. Moreover, we will denote with $t$ the time variable.
We recall now the governing equations for capillary-gravity water waves arising as the free surface (denoted $y=\eta(x,t)$) of a two dimensional rotational water flow of constant vorticity bounded below by the flat bed $y=-h$ ($h$ being some positive constant) and above by the surface itself, see Fig. \ref{fig1}.

Denoting with $(u(x,y,t),v(x,y,t))$ the velocity field, with $P(x,y,t)$ the pressure and with $g$ the gravitational constant
the equations of motion are the Euler equations

\begin{equation}\label{Eulereq}
\begin{split}
u_t+uu_x+vu_y&=-P_x\\
v_t+uv_x+vv_y&=-P_y-g,
\end{split}
\end{equation}
and the equation of mass conservation
\begin{equation}\label{masscons}
u_x+v_y=0.
\end{equation}
The systems \eqref{Eulereq} and \eqref{masscons} are completed by the kinematic boundary conditions
\begin{equation}\label{BKS}
\begin{split}
v&=\eta_t + u \eta_x\quad{\rm on}\quad y=\eta(x,t),\\
v&=0\quad{\rm on}\quad y=-h,
\end{split}
\end{equation}
and by the dynamic condition
$$
P=P_{\rm atm}-\tilde{\sigma} \f{\eta_{xx}}{(1+\eta_x^2)^{\f{3}{2}}}\quad{\rm on}\quad y=\eta(x,t),
$$
where $\tilde{\sigma}$ denotes the coefficient of surface tension.

The fluid motion in the presence of a steady current can be written by means of a (generalized) velocity potential $\varphi$
as
\begin{equation}
u=\varphi_x(x,y,t)+\gamma y + \kappa, \qquad v=\varphi_y(x,y,t),
\end{equation} where $\kappa$ represents the average constant velocity of the current and $\gamma$ is a constant
which represents the constant vorticity of the flow since the curl of the velocity field is
\begin{equation}\label{vort}
u_y-v_x=\gamma\,\,{\rm within}\,\,{\rm the}\,\,{\rm flow}.
\end{equation}

For simplicity we assume
$\kappa=0$, a scenario that is attained by a Galilean coordinate transformation and a proper choice of a reference frame.

Setting $\Phi(x,t):=\varphi(x,\eta(x,t),t)$, the one-dimensional wave propagation in the $x$ direction can be expressed in a {\it nearly}-Hamiltonian form  \cite{NearlyHamiltonian}
\begin{equation}\label{NH}
\f{\delta H}{\delta\Phi}=\eta_t \quad {\rm and}\quad -\f{\delta H}{\delta\eta}=\Phi_t +\gamma\partial_x^{-1}\eta_t.
\end{equation} where $H[\Phi,\eta]=\tilde{H}/\rho$, where $\rho$ is the fluid density and $\tilde{H}$ is the total energy of the fluid in terms of $\Phi, \eta.$

\subsection{The Hamiltonian for the capillary-gravity waves}

In the case of capillary-gravity waves the Hamiltonian functional is
\begin{equation}
\tilde{ H}=\frac{\rho}{2}\int \Phi G(\eta)\Phi\,dx+\frac{\gamma\rho}{2}\int\Phi_x\eta^2\,dx+\frac{\gamma^2\rho}{6}\int\eta^3\,dx+\frac{g\rho}{2}\int\eta^2\,dx+\tilde{\sigma}\int\sqrt{1+\eta_x^2}\,dx,
\end{equation} where $G(\eta)$ is the so-called Dirichlet-Neumann operator \cite{CraigGroves1} and $\tilde{\sigma}$ is the surface tension coefficient. Introducing
$$ \sigma:= \tilde{\sigma} / \rho $$ the Hamiltonian is

\begin{equation}
H=\frac{1}{2}\int \Phi G(\eta)\Phi\,dx+\frac{\gamma}{2}\int\Phi_x\eta^2\,dx+\frac{\gamma^2}{6}\int\eta^3\,dx+\frac{g}{2}\int\eta^2\,dx+\sigma \int\sqrt{1+\eta_x^2}\,dx
\end{equation}

Next, we need to introduce some relevant scales and to specify the order of magnitude of the physical quantities under investigation. We introduce, as usual, the scale parameters $$\varepsilon:=\frac{a}{h} \quad \text{and} \quad \delta:= \frac{h}{\lambda}$$ where $a$ is the wave amplitude ($0<|\eta(x,t)|\le a$) and $\lambda$ is the wavelength, and consider a shallow water scaling regime with $\delta \simeq 1$ and $\varepsilon \ll 1$.

The Dirichlet-Neumann operator is
$$G=G_0+G_1+G_2+\ldots$$ where $G_k(\eta)$ is a operator, such that $G_k(\mu\eta)=\mu^kG_k(\eta)$ for any constant $\mu$; thus, given that $\eta \simeq \varepsilon h$ is of order $\varepsilon$, $G_k(\eta)$ is of order $\varepsilon^k$. More specifically,
$G_0=D\tanh(Dh)=DT$, with $D=-i\partial_x$ and $T:=\tanh(Dh)$.

Given that the operator $hD$ has a Fourier multiplier $hk\simeq 2 \pi h/ \lambda \simeq 2 \pi $  as far as $\delta \simeq 1$, we have
$|T|=|\tanh(2\pi)|\approx 1$. Hence, $T\rightarrow \sign (D)$ and we have $G_0\rightarrow |D|\simeq \mathcal{O}(1)$.
Moreover, $ \partial_x$ is of the order of $ (1/h) (hD)\simeq 2\pi/h$ which is a fixed constant and therefore is $ \mathcal{O}(1) .$ The time derivative is assumed to be also of order $1$.
Noticing that the gradient of $\Phi$ has a magnitude of a velocity, and given \eqref{BKS}, $\Phi$ is of the same order as $\eta$, i.e. of order $\varepsilon$.

\begin{gather*}
 G_1=D\eta D-DT\eta DT\rightarrow D\eta D-|D|\eta |D|\simeq \mathcal{O}(\varepsilon)\\
 G_2\simeq\mathcal{O}(\varepsilon^2), \quad \text{etc.} \\
 \eta \simeq \mathcal{O}(\varepsilon), \qquad \Phi \simeq\mathcal{O}(\varepsilon).
\end{gather*}
 Therefore, writing explicitly the scale parameter, and keeping terms up to $\varepsilon^3$ we obtain
\begin{align*}
 H(\eta,\Phi)=&\f{\varepsilon^2}{2}\int\Phi |D|\Phi\,dx+\f{\varepsilon^2 g}{2}\int\eta^2 \,dx+\f{\varepsilon^2 \sigma}{2}\int\eta_x^2\,dx\\
 &+\f{\gamma\e^3}{2}\int\Phi_x\eta^2\,dx+\f{\gamma^2\e^3}{6}\int\eta^3\,dx+\f{\e^3}{2}\int\Phi_x\eta\Phi_x\,dx-\f{\e^3}{2}\int (|D|\Phi)\eta (|D|\Phi)\, dx
\end{align*} The meaning of the operator $|D|$ is explained in the Appendix.

Let us denote with $H_0$ the leading order part of the Hamiltonian, that is
\begin{equation}\label{H0}
H_0=\f{1}{2}\int\Phi |D|\Phi\,dx+\f{ g}{2}\int\eta^2 \,dx+\f{ \sigma}{2}\int\eta_x^2\,dx.
\end{equation}
On the other hand, from \eqref{H0} and \eqref{NH} we have
\begin{equation}
\f{\delta H_0}{\delta\Phi}=|D|\Phi\quad {\rm and}\quad \f{\delta H_0}{\delta\eta}=g\eta-\sigma\eta_{xx}.
\end{equation}
Hence, from the previous two systems of equations we derive the system
\begin{equation}
\begin{aligned}
\eta_t=|D|\Phi\\
\Phi_t +\gamma\partial_x^{-1}\eta_t=-g\eta+\sigma\eta_{xx}.
\end{aligned}
\end{equation}
Applying now the operator $|D|$ to the second equation of the above system and taking the first one into account we can eliminate $\Phi$ and obtain the equation for $\eta$
\begin{equation}\label{eta}
\eta_{tt}-i\gamma\sign(D)\eta_t +g|D|\eta-\sigma |D|\eta_{xx}=0.
\end{equation}
Looking for solutions of the type $\eta=\eta_0 e^{kx-\omega(k)t}$ of equation \eqref{eta} we obtain that the frequency $\omega(k)$ satisfies the equation
\begin{equation}\label{disp_eq}
\omega^2+\gamma\sign(k)\omega-|k|(g+\sigma k^2)=0,
\end{equation}
whose solutions are given by the dispersion relation
\begin{equation}\label{frequency_form}
\omega(k)=\f{-\gamma\sign(k)\pm\sqrt{\gamma^2+4|k|(g+\sigma k^2)}}{2}.
\end{equation}

For purely gravity waves the term containing $g$ dominates, i.e. $\omega(k) \simeq \sqrt{|k|g}$ while for mostly capillary waves $\omega(k)\simeq \sqrt{\sigma |k^3|}=\sqrt{\frac{\tilde{\sigma} | k^3|}{\rho}}$. The balance between capillary and gravity effects occurs for $gk\simeq \frac{\tilde{\sigma}  k^3}{\rho}$, expression which in terms of wavelengths
becomes
\begin{equation}\label{est} \lambda = \frac{2\pi}{k} \simeq 2\pi \sqrt{\frac{\tilde{\sigma}}{\rho g}}.\end{equation}

The latter expression implies that for capillary-gravity water waves we can confidently work with values of the wavelength that revolve around $\lambda \simeq 17$ mm.

\section{Resonance conditions for the capillary-gravity waves}
We recall in this section the condition for three-wave resonance to occur as well as the expression of the vorticity that
gives rise to resonance, cf. \cite{CK, MarJmfm}. The occurrence of three-wave resonance is of capital  importance for the computation of the cubic terms in the Hamiltonian
function in Section \ref{waveampl_eqs}.

Formula \eqref{frequency_form} is of interest for the investigation of three-wave resonances. This is equivalent to the
existence of wave numbers $k_1, k_2$ and $k_3$ that satisfy the system
\begin{equation} \label{res_syst}
k_1+k_2=k_3, \qquad \omega(k_1)+\omega(k_2)=\omega(k_3)
\end{equation}

Three-wave resonance is a highly significant process that occurs between two waves that combine to give rise to a new, third wave. While three-wave resonances between surface gravity waves are not possible, the presence of a not to small
constant vorticity does trigger the emergence of resonances in the case of capillary \cite{CK} and of capillary-gravity water waves \cite{MarJmfm}. More precisely, a triple of integers $(k_1,k_2,k_3)$ solves \eqref{res_syst} if and only
if
\begin{equation}\label{cond_res}
\sqrt{k_3^3+k_3\f{g}{\sigma}}>\sqrt{k_1^3+k_1\f{g}{\sigma}}+\sqrt{k_2^3+k_2\f{g}{\sigma}}.
\end{equation}
If \eqref{res_syst} holds, then the vorticity $\gamma$ (assuming positive  vorticity $\gamma$) could be determined if the resonance triple $k_1, k_2, k_3$
is known by the formula
\begin{equation}
\gamma = \sqrt{\sigma} \f{|\mathcal{E}^2-4k_1^3 k_2^3-A|}{4 \mathcal{E}\left(\mathcal{E}^2-4k_1^3 k_2^3+2k_3^3 \mathcal{E}+2k_3 \f{g}{\sigma} \mathcal{E}-A\right) }
\end{equation}
where $\mathcal{E}=k_1^3+k_2^3-k_3^3$ and $A=4k_1k_2\f{g}{\sigma}\left(\f{g}{\sigma}+k_1^2+k_2^2\right)$.
\section{Derivation of equations for wave amplitudes}\label{waveampl_eqs}

In order to study the propagation and the interaction of wave packages, for each wave package (numbered by $j$) we introduce the ``fast'' oscillating carrier wave with a fixed wavelength,   $E_j=e^{i(k_j x-\omega_j t)},\,\omega_j=\omega(k_j)$ and a ``slow'' varying wave envelope $a_j=a_j(X,T)$.  The ``slow'' variables depend on $X:=\varepsilon x \simeq \mathcal{O} (\varepsilon)$ and $T:=\varepsilon t \simeq \mathcal{O} (\varepsilon).$  The wave packages are assumed to be well separated, so that we assume also $$ \lim_{X\to \pm \infty}a_j(X,T)=0.$$

 We consider several interacting wave packages represented in the form \cite{Cra}
\begin{gather}
\eta(x,t)=\sum_j M_j (a_j E_j+\ov{a_j}\ov{E_j})\\
\Phi(x,t)=\sum_j -i N_j(a_j E_j-\ov{a_j}\ov{E_j})
\end{gather} where $M_j$ and $N_j$ are constants of orders $ \simeq \mathcal{O} (\varepsilon)$ while the variables $a_j$ and $E_j$ are of order 1.

We proceed now to evaluate explicitly the terms appearing in the Hamiltonian by means of the building blocks $a_j$ and $E_j$.
\begin{align*}
\partial_x^{-1}(a(X,T)E)&=\int_{-\infty}^x a(\tilde{X},T)e^{i(k\tilde{x}-\omega t)}\,d\tilde{x}=\int_{-\infty}^x a(\tilde{X},T)\f{\partial}{\partial \tilde{x}}\left(\f{e^{i(k\tilde{x}-\omega t)}}{ik}\right)\,d\tilde{x}\\
&=a(\tilde{X},T)\f{e^{i(k\tilde{x}-\omega t)}}{ik}\Big|_{\tilde{x}=-\infty}^{\tilde{x}=x}-\f{1}{ik}\int_{-\infty}^x e^{i(k\tilde{x}-\omega t)}\e a^{\prime}(\tilde{X},T)\,d\tilde{x}\\
&=\f{a E}{ik}-\f{\e}{ik} \partial_x^{-1}(a'E)
\end{align*} where the obvious notation $E=\exp(i(kx-\omega t))$ has been used and the prime denotes a derivative with respect to the $X$ variable. By repeating the previous procedure and taking into account the considerations in Section \ref{slowafast} we find that
\begin{equation}
\partial_x^{-1}(a_j E_j)=\f{1}{ik_j}a_j E_j-\f{\e}{(ik_j)^2}a^{\prime}_j E_j+\f{\e^2}{(ik_j)^3}a_j^{\prime\prime} E_j+ \mathcal{O}(\varepsilon^3)
\end{equation}
Likewise
\begin{align}
\partial_x^{-1}\eta_t =&\partial_x^{-1}\sum_{j} [M_j(\e\dot{a}_j-i\omega_j a_j)E_j +M_j( \e \dot{\ov{a}}_j+i\omega_j\ov{a_j})\ov{E_j}  ] \nonumber \\
=& \sum_{j} M_j[\e\partial_x^{-1}(\dot{a}_j E_j+\dot{\ov{a}}_j\ov{E_j})  -i\omega_j \partial_x^{-1}(a_j E_j-\ov{a_j}\ov{E_j})] \nonumber \\
=& \sum_{j} M_j [ \f{\e}{i k_j}(\dot{a}_j E_j-\dot{\ov{a}}_j\ov{E_j})-\f{\e^2}{(ik_j)^2}(\dot{a}_j ^{\prime} E_j+\dot{\ov{a}}_j^{\prime}\ov{E_j})-\f{i\omega_j}{i k_j}(a_j E_j+\ov{a_j}\ov{E_j})
\nonumber \\
 &+\f{\e i\omega_j}{(i k_j)^2}(a_j^{\prime} E_j-\ov{a_j}^{\prime}\ov{E_j})-\f{\e^2 i\omega_j}{(i k_j)^3}(a_j^{\prime\prime} E_j+\ov{a_j}^{\prime\prime}\ov{E_j})]
 + \mathcal{O}(\varepsilon^3)\nonumber
\end{align}

Thus
\begin{align}
\partial_x^{-1}\eta_t =\sum_{j}M_j\left[-\f{\omega_j}{k_j}(a_j E_j+\ov{a}_j\ov{E}_j)-\f{\e i\omega_j}{k_j^2}(a_j^{\prime}E_j-\ov{a}_j^{\prime}\ov{E}_j)-\f{\e i}{k_j}(\dot{a}_j E_j-\dot{\ov{a}}_j \ov{E}_j)\right]+\mathcal{O}(\e^2)
\end{align}
We use now the (nearly) Hamiltonian formulation \eqref{NH} and write
\begin{align*}
\delta H=&\eta_t\delta\Phi-(\Phi_t+\gamma\partial_x^{-1}\eta_t)\delta\eta\\
=&\sum_{j,p}\int[M_j(\e\dot{a}_j-i\omega_j a_j)E_j+M_j(\e\dot{\ov{a}}_j+i\omega_j\ov{a}_j\ov{E}_j)][-iN_p][E_p\delta  a_p-\ov{E}_p\delta \ov{a}_p]\,dx\\
&- \sum_{j,p} \int(-iN_p)\Big[ \big( (\e \dot{a}_p-i\omega_p a_p)E_p-(\e\dot{\ov{a}}_p+i\omega_p\ov{a}_p)\ov{E}_p\big)\\
&+\gamma M_p\Big(-\f{\omega_p}{k_p}(a_p E_p+\ov{a}_p\ov{E}_p)-\e\f{i\omega_p}{k_p^2}(a^{\prime}_p E_p-\ov{a}^{\prime}_p \ov{E}_p)-\e\f{i}{k_p}(\dot{a}_p E_p-\dot{\ov{a}}_p\ov{E}_p)\Big)\Big]\\
&\times M_j (E_j\delta a_j+\ov{E}_j\delta\ov{a}_j)\,dx + \mathcal{O}(\varepsilon ^4)
\end{align*}
If not specified, the integration is over the entire real axis. Since integration of products between ``slow'' and ``fast'' terms vanishes (see the Appendix) we are left with the following

\begin{align}
\delta H=&2 \sum_{j} \int iM_jN_j\big[(\e\dot{a}_j-i\omega_j a_j)\delta\ov{a}_j-(\e\dot{\ov{a}}_j+i\omega_j \ov{a}_j)\delta a_j\big]\,dx\\
&+\gamma \sum_{j}\int M_j^2\left(\f{\omega_j}{k_j} a_j+\e\f{i\omega_j}{k_j^2}a^{\prime}_j+\e\f{i}{k_j}\dot{a}_j\right)\delta\ov{a}_j\,dx\\
&+\gamma \sum_{j} \int M_j^2\left(\f{\omega_j}{k_j} \ov{a}_j-\e\f{i\omega_j}{k_j^2}\ov{a}^{\prime}_j-\e\f{i}{k_j}\dot{\ov{a}}_j\right)\delta a_j\,dx
+ \mathcal{O}(\varepsilon ^4)
\end{align}
We then obtain
\begin{equation}\label{firstformvar}
\f{\delta H}{\delta \ov{a}_j}=i\left(2M_j N_j+\f{\gamma M_j^2}{k_j}\right)(\e\dot{a}_j-i\omega_j a_j)+\e \gamma M_j^2\f{i\omega_j}{k_j^2}a^{\prime}_j +\mathcal{O}(\e^4)
\end{equation}
We calculate now the first term of $H_0$ in \eqref{H0}. By taking into account the properties of the Hilbert transform (see the Appendix) we obtain

\begin{align} \label{phidphi}
\f{1}{2}&\int \Phi |D|\Phi\,dx=\nonumber  \\
=& \f{1}{2} \sum_{j,p} \int(-i N_j)(a_j E_j-\ov{a}_j \ov{E}_j)(\mathcal{H}\partial_x)(-i N_p)(a_p E_p-\ov{a}_p \ov{E}_p)\,dx \nonumber \\
=&-\f{1}{2} \sum_{j,p} N_j N_p \int (a_j E_j-\ov{a}_j \ov{E}_j)\big[(\e a_p^{\prime}\mathcal{H}+a_p|k_p|)E_p-(\e \ov{a}_p^{\prime}\mathcal{H}+\ov{a}_p|k_p|)\ov{E}_p\big]\,dx \nonumber  \\
=&-\f{1}{2} \sum_{j} N_j^2\int\big[-a_j(\e\ov{a}_j^{\prime}i\sign(k_j)+\ov{a}_j|k_j|)-\ov{a}_j(\e a_j^{\prime}(-i)\sign(k_j)+a_j|k_j|)\big]\,dx  \nonumber \\
=& \sum_{j} N_j^2\int a_j\ov{a}_j|k_j|\,dx+\e \sum_{j} \f{i N_j^2}{2}\int\sign(k_j)(a_j\ov{a}_j^{\prime}-a_j^{\prime}\ov{a}_j)\,dx .
\end{align}
Similarly,
\begin{equation}\label{g}
\f{g}{2} \int \eta^2(x)\,dx=g \sum_{j} M_j^2\int a_j\ov{a}_j\,dx ,
\end{equation}
and
\begin{align}
\f{\sigma}{2}\int\eta_x^2\,dx & = \f{\sigma}{2} \sum_{j,p} \int\hspace{-0.15cm} M_j\big[(\e a_j^{\prime}+i k_j a_j)E_j+(\e\ov{a}_j^{\prime}-i k_j\ov{a}_j)\ov{E}_j\big]
\nonumber \\
&\hspace{2cm} \times M_p\big[(\e a_p^{\prime}+i k_p a_p)E_p+(\e\ov{a}_p^{\prime}-i k_p\ov{a}_p)\ov{E}_p\big] \, dx  \nonumber \\
&\hspace{-0.3cm}=\sum_{j} \sigma M_j^2\int\big[\e^2 a_j^{\prime}\ov{a}_j^{\prime}-\e i k_j(\ov{a}_j a_j^{\prime}-\ov{a}_j^{\prime}a_j)+k_j^2 a_j\ov{a}_j\big]\,dx
\label{sigma}.
\end{align}
Considering all the relevant contributions from \eqref{phidphi},\eqref{g} and  \eqref{sigma} we have
\begin{equation}\label{secondformvar}
\f{\delta H_0}{\delta\ov{a}_j}=N_j^2 |k_j| a_j+g M_j^2 a_j +\sigma M_j^2 k_j^2 a_j-\e a_j^{\prime}\big[i\sign(k_j)N_j^2+2 i\sigma M_j^2 k_j\big]
\end{equation}
Comparing now the coefficients of $\e^2$ from \eqref{firstformvar} and \eqref{secondformvar} we obtain the equality
\begin{equation}
\left(2M_j N_j+\gamma\f{M_j^2}{k_j}\right)\omega_j a_j=N_j^2 |k_j| a_j+g M_j^2 a_j +\sigma M_j^2 k_j^2 a_j,
\end{equation}
which leads to the following quadratic equation for the unknown $\f{N_j}{M_j}$,

\begin{equation}
\left(\f{N_j}{M_j}\right)^2|k_j|-2\omega_j \left(\f{N_j}{M_j}\right)-\f{\gamma\omega_j}{k_j}+g+\sigma k_j^2=0
\end{equation}
with solutions
\begin{equation}
\f{N_j}{M_j}=\f{\omega_j}{|k_j|}\pm\f{1}{|k_j|}\sqrt{\omega_j^2+\sign(k_j)\gamma\omega_j-|k_j|(g+\sigma k_j^2)}=\f{\omega_j}{|k_j|},
\end{equation}
by the virtue of \eqref{disp_eq}.

From \eqref{firstformvar} we have now

\begin{equation}\label{firstformvar1}
\f{\delta H_r}{\delta \ov{a}_j}=\varepsilon i\left[\left(2M_j N_j+\f{\gamma M_j^2}{k_j}\right)\dot{a}_j + \gamma M_j^2\f{\omega_j}{k_j^2}a^{\prime}_j\right] +\mathcal{O}(\e^4)
\end{equation}

where $H_r=H_{01}+H_1$ contains the $\mathcal{O}(\varepsilon^3) $ terms from $H_0$,

$$H_{01}=  \e i \sum_{j} \left(\f{ N_j^2}{2} \sign(k_j) + \sigma M_j^2 k_j \right) \int (a_j\ov{a}_j^{\prime}-a_j^{\prime}\ov{a}_j)\,dx  $$

and the contribution from the cubic terms $H_1$ which will be evaluated later.

The order $\e^3$ terms in the equation \eqref{firstformvar1} yield the equality
\begin{equation}
\dot{a}_j+\f{ \left(\f{N_j}{M_j}\right)^2\sign (k_j)+\left(2\sigma k_j+\gamma\f{\omega_j}{k_j^2}\right)}{2\f{N_j}{M_j}+\f{\gamma}{k_j}}a_j^{\prime}=
\frac{1}{i \varepsilon \left( 2M_j N_j + \gamma (M_j^2/k_j) \right)} \frac{\delta H_1}{\delta \ov{a}_j},
\end{equation}
which, with the choice $\f{N_j}{M_j}=\f{\omega_j}{|k_j|}$ becomes
\begin{equation}
\dot{a}_j+\f{ \left(\f{\omega_j}{|k_j|}\right)^2\sign (k_j)+\left(2\sigma k_j+\gamma\f{\omega_j}{k_j^2}\right)}{2\f{\omega_j}{|k_j|}+\f{\gamma}{k_j}}a_j^{\prime}=\frac{1}{i \varepsilon \left( 2M_j N_j + \gamma (M_j^2/k_j) \right)} \frac{\delta H_1}{\delta \ov{a}_j}.
\end{equation}
We set now
\begin{equation}\label{vj}
v_j:=\f{ \left(\f{\omega_j}{|k_j|}\right)^2\sign (k_j)+\left(2\sigma k_j+\gamma\f{\omega_j}{k_j^2}\right)}{2\f{\omega_j}{|k_j|}+\f{\gamma}{k_j}}=\f{\sign(k_j)\omega_j^2+2\sigma k_j^3+\gamma\omega_j}
{k_j(2\omega_j \sign(k_j)+\gamma)}.
\end{equation}
We will prove in the sequel that  $v_j$ defined in \eqref{vj} equals, in fact, the group velocity $\f{\p\omega}{\p k}$. Indeed, differentiating in \eqref{disp_eq}, we find first that
\begin{equation}
2\omega\f{\p\omega}{\p k}+\sign(k)\gamma\f{\p\omega}{\p k}-(g+3\sigma k^2)\sign(k)=0,
\end{equation}
which yields
\begin{align*}
\f{\p\omega}{\p k}=&\f { (g+3\sigma k^2)\sign(k)}{2\omega +\gamma\sign(k)}=\f { (g+3\sigma k^2)|k|}{k[2\omega +\gamma\sign(k)]}=\f{\omega^2+\gamma\sign(k)\omega+2\sigma k^2 |k|}{k[2\omega +\gamma\sign(k)]}\\
=&\f{\sign(k)\omega^2+\gamma \omega+2\sigma k^3}{k(2\omega\sign(k)+\gamma)}.
\end{align*}
The latter equality implies that $\f{\p\omega}{\p k}(k_j)=v_j$.

Moreover, from the previous considerations, it emerges that
\begin{equation}\label{RS}
\dot{a}_j+v_j a_j^{\prime}=\frac{1}{i \varepsilon \left( 2M_j N_j + \gamma (M_j^2/k_j) \right)} \frac{\delta H_1}{\delta \ov{a}_j}.
\end{equation}

Next, we compute the contribution from the cubic terms in the Hamiltonian,
\begin{equation}
H_1= \f{1}{2}\int\Phi_x\eta\Phi_x\,dx-\f{1}{2}\int (|D|\Phi)\eta (|D|\Phi)\,dx+\f{\gamma}{2}\int\Phi_x\eta^2\,dx+\f{\gamma^2}{6}\int\eta^3\,dx.
\end{equation}
Let us assume a resonant interaction between $3$ wave packages as in \eqref{res_syst}, such that $k_3=k_1+k_2$. Without loss of generality we can assume all $k_n>0$ then
\begin{equation}
\begin{split}
H_{11}&=\f{1}{2}   \int\Phi_x\eta\Phi_x\,dx\\
&=\f{1}{2}  \sum_{j,p,n} \int \Big\{  (-i N_j)[(\e a_j^{\prime}+i k_j a_j) E_j-(\e \ov{a}_j^{\prime} -ik_j\ov{a}_j )\ov{E}_j]M_p(a_p E_p+\ov{a}_p\ov{E}_p)\Big. \\
 &\hspace{2cm }\Big. \times (-i N_n)[(\e a_n^{\prime}+i k_n a_n) E_n-(\e \ov{a}_n^{\prime} -ik_n\ov{a}_n)\ov{E}_n]\Big\}\,dx\\
&=\f{1}{2}\sum_{j,p,n} \int N_j M_p N_n k_j k_n (a_j E_j+\ov{a}_j \ov{E}_j) (a_p E_p+\ov{a}_p\ov{E}_p) (a_n E_n+\ov{a}_n \ov{E}_n),dx
\end{split}
\end{equation}
Since the indices $j,p,n$ run from $1$ to $3$ each and due to \eqref{res_syst} we obtain
\begin{equation}
\begin{split}
H_{11}=&M_1 N_2 N_3 k_2 k_3\int (a_1 a_2\ov{a}_3+\ov{a}_1\ov{a}_2 a_3)\,dx\\
&+M_2 N_1 N_3 k_1 k_3 \int (a_1 a_2\ov{a}_3+\ov{a}_1 \ov{a}_2 a_3)\,dx\\
&+M_3 N_1 N_2 k_1 k_2\int (a_1 a_2\ov{a}_3+\ov{a}_1 \ov{a}_2 a_3)\,dx .
\end{split}
\end{equation}
Without loss of generality we make the choice $M_1=M_2=M_3=\e$ which yields $N_j=\e\f{\omega_j}{k_j}$. Therefore,
\begin{equation}
H_{11}=\e^3(\omega_1\omega_2+\omega_2\omega_3+\omega_3\omega_1)\int (a_1 a_2\ov{a}_3+\ov{a}_1 \ov{a}_2 a_3)\,dx.
\end{equation}
We compute now $H_{12}=-\f{1}{2}\int (|D|\Phi)\eta (|D|\Phi)\,dx$. To this end note that
$$|D|(a_j E_j)=(\partial_x\mathcal{H})(a_j E_j)=\mathcal{H}(\e a^{\prime}_j E_j+a_j i k_j E_j)=ik_j\mathcal{H}( a_j E_j)+\mathcal{O}(\e).$$
Since the Fourier spectra of the slow $a_j$ and of the fast $E_j$ do not overlap, using the result from \cite{Bed} we have $\mathcal{H}(a_j E_j)=a_j \mathcal{H}E_j$. Hence,
\begin{equation}\label{absDaj}
|D|(a_j E_j)=|k_j|a_j E_j+\mathcal{O}(\e).
\end{equation}
Similarly to the above computations we have
\begin{equation}\label{absDbaraj}
|D|(\ov{a}_j \ov{E}_j)=|k_j|\ov{a}_j \ov{E}_j+\mathcal{O}(\e).
\end{equation}
Using \eqref{absDaj} and \eqref{absDbaraj} we compute
\begin{equation}
\begin{split}
H_{12}=&\f{1}{2} \sum_{j,p,n}\int N_j M_p N_n |k_j| |k_n|(a_j E_j-\ov{a}_j\ov{E}_j)(a_p E_p+\ov{a}_p\ov{E}_p)(a_n E_n-\ov{a}_n\ov{E}_n)\,dx\\
=&M_1 N_2 N_3 |k_2| |k_3|\int(a_1 a_2 (-\ov{a}_3)+\ov{a}_1 (-\ov{a}_2) a_3)\,dx\\
&+M_2 N_1 N_3 |k_1| |k_3|\int(a_1 a_2 (-\ov{a}_3)+(-\ov{a}_1)\ov{a}_2 a_3)\,dx\\
&+M_3 N_1 N_2 |k_1| |k_2|\int(a_1 a_2\ov{a}_3+(-\ov{a}_1)(-\ov{a}_2) a_3)\,dx\\
=&\e^3 (-\omega_2\omega_3 -\omega_1\omega_3+\omega_1\omega_2)\int(a_1 a_2\ov{a}_3+\ov{a}_1\ov{a}_2 a_3)\,dx
\end{split}
\end{equation}
Further, we compute
\begin{equation}
\begin{split}
H_{13}=&\f{\gamma^2}{6}\int\eta^3\,dx\\
=&\f{\gamma^2}{6} \sum_{j,p,n}  M_j M_p M_n\int (a_j E_j+\ov{a}_j \ov{E}_j)(a_p E_p+\ov{a}_p\ov{E}_p)(a_n E_n+\ov{a}_n\ov{E}_n)\,dx\\
=&6\cdot\f{\gamma^2}{6} M_1 M_2 M_3\int(a_1 a_2\ov{a}_3+\ov{a}_1 \ov{a}_2 a_3)\,dx\\
=&\e^3 \gamma^2\int(a_1 a_2\ov{a}_3+\ov{a}_1 \ov{a}_2 a_3)\,dx.
\end{split}
\end{equation}
Finally,
\begin{equation}
\begin{split}
H_{14}=&\f{\gamma}{2}\int\Phi_x\eta^2\,dx\\
=&\f{\gamma}{2} \sum_{j,p,n} (-iN_j)(i k_j)\int(a_j E_j+\ov{a}_j\ov{E}_j)M_p(a_p E_p+\ov{a}_p \ov{E}_p)M_n(a_n E_n+\ov{a}_n \ov{E}_n)\, dx\\
=&\f{\gamma}{2} \cdot 2 N_1 k_1 M_2 M_3\int(a_1 a_2\ov{a}_3+\ov{a}_1 \ov{a}_2 a_3)\,dx\\
&+\f{\gamma}{2}\cdot 2 N_2 k_2 M_1 M_3\int(a_1 a_2\ov{a}_3+\ov{a}_1 \ov{a}_2 a_3)\,dx\\
&+\f{\gamma}{2} \cdot 2 N_3 k_3 M_1 M_2\int(a_1 a_2\ov{a}_3+\ov{a}_1 \ov{a}_2 a_3)\,dx\\
=&\e^3\gamma (\omega_1 +\omega_2 +\omega_3)\int(a_1 a_2\ov{a}_3+\ov{a}_1 \ov{a}_2 a_3)\,dx.
\end{split}
\end{equation}
Counting all the contributions computed before we get
\begin{equation}
\begin{split}
H_1=&\e^3[\gamma^2+\gamma(\omega_1+\omega_2+\omega_3)+2\omega_1\omega_2]\int(a_1 a_2\ov{a}_3+\ov{a}_1 \ov{a}_2 a_3)\,dx\\
=&\e^3(\gamma^2+2\omega_3\gamma+2\omega_1\omega_2)\int(a_1 a_2\ov{a}_3+\ov{a}_1 \ov{a}_2 a_3)\,dx,
\end{split}
\end{equation}
where, for the second equality, we have used the resonance relation $\omega_3=\omega_1+\omega_2$.
Taking \eqref{RS} into account we have
\begin{equation}\label{ampl_eqs}
\begin{split}
\dot{a}_1+v_1 a_1^{\prime}=-i k_1\f{\gamma^2 +2\omega_3\gamma +2\omega_1\omega_2}{2\omega_1+\gamma}\ov{a}_2 a_3\\
\dot{a}_2+v_2 a_2^{\prime}=-i k_2\f{\gamma^2 +2\omega_3\gamma +2\omega_1\omega_2}{2\omega_2+\gamma}\ov{a}_1 a_3\\
\dot{a}_3+v_3 a_3^{\prime}=- i k_3\f{\gamma^2 +2\omega_3\gamma +2\omega_1\omega_2}{2\omega_3+\gamma}a_1 a_2
\end{split}
\end{equation}
Setting $\gamma=0$ in \eqref{ampl_eqs} we recover the system of equations for the irrotational case
\begin{equation}
\begin{split}
\dot{a}_1+v_1 a_1^{\prime}=-i k_1\omega_2\ov{a}_2 a_3\\
\dot{a}_2+v_2 a_2^{\prime}=-i k_2\omega_1\ov{a}_1 a_3\\
\dot{a}_3+v_3 a_3^{\prime}=-i k_3\f{\omega_1\omega_2}{\omega_3} a_1 a_2
\end{split}
\end{equation}
cf. equation (14.8) from \cite{Cra} where the work of Case and Chiu \cite{CaCh} is outlined.
We observe also that there are two special values for the vorticity, $$\gamma^*=\omega_1+\omega_2 \pm \sqrt{\omega_1^2+\omega_2^2} $$ for which $H_1=0$
and the resonance is destroyed.

The system \eqref{ampl_eqs} is integrable, it could be brought to the standard form \eqref{3wso} from the Appendix by a rescaling
$a_j \rightarrow \mu_j u_j$ for the constants

$$\mu_j=  \frac{-1}{\gamma^2 +2\omega_3\gamma +2\omega_1\omega_2}\sqrt{\frac{\eta_j}{\eta_1 \eta_2 \eta_3}}, \qquad \eta_j=\frac{k_j}{2\omega_j + \gamma}. $$

From the system \eqref{ampl_eqs} it is evident that the variables $a_1$ and $a_2$ are entering in a symmetric way while the $a_3$ variable is a different one. This is due to the physical interpretation of the system: The complex $a_j$ variables describe the envelopes of three wave packages, where $a_1$ and $a_2$ can merge into $a_3$ or $a_3$ can decay into $a_1$ and $a_2$, see for example \cite{ZMNP,ZM} for details.

Assuming further that all $\omega_n>0$ and $\gamma>0$ we evaluate
\begin{equation}
v_3=\f{\omega_3^2+\gamma\omega_3+2\sigma k_3^3}{k_3(2\omega_3+\gamma)}=
\f{k_3(g+\sigma k_3^2)+2\sigma k_3^3}{k_3(2\omega_3+\gamma)}=\f{g+3\sigma k_3^2}{2\omega_3+\gamma},
\end{equation}
where the second equality sign holds because of the equation satisfied by $\omega$. Clearly, similar formulas are verified by $v_1$ and $v_2$, thus all $v_n$ are positive. We will demonstrate that $v_3$ is the largest among the three group velocities. To this end we compute $v_3^2$. In doing so we will use
that $(2\omega_3+\gamma)^2=\gamma^2+4k(g+\sigma k^2)$ as it follows from formula (2.14) when we assume $\sign(k)=1$. Hence, we have
\begin{equation}
v_3^2=\f{(g+3\sigma k_3^2)^2}{\gamma^2+4k_3(g+\sigma k_3^2)},
\end{equation}
and, of course, similar formulas for $v_1^2$ and $v_2^2$ hold as well.

Let us introduce the function
\begin{equation}
f(k)=\f{(g+3\sigma k^2)^2}{\gamma^2+4k(g+\sigma k^2)}.
\end{equation}
Computing the derivative of $f$ we find that the numerator of the arising fraction is
\begin{equation}\label{numerator}
\begin{split}
&4k(g+3\sigma k^2)^2\cdot 3\sigma(\gamma^2+4k(g+\sigma k^2)) -(g+3\sigma k^2)^2\cdot 4(g+3\sigma k^2)\\
=&4(g+3\sigma k^2)^2 [3\sigma\gamma^2 k+12\sigma g k^2+12\sigma^2 k^4-g^2-6\sigma g k^2-9\sigma^2 k^4]\\
=&4(g+3\sigma k^2)^2[3\sigma k (\gamma^2+2g k+\sigma k^3)-g^2].
\end{split}
\end{equation}
We note that for capillary-gravity waves with wavelength smaller or equal to $2$ cm, we have that
$2\sigma k^2$ exceeds $g$. Thus, $6\sigma g k^2>3g^2$, which implies that the second bracket in \eqref{numerator}
is positive. This shows that $f(k)$ is an increasing function of $k$.
Since $k_3>k_1$ and $k_3>k_2$ we have $v_3^2>v_1^2$ and $v_3^2>v_1^2$, that is $v_3>v_1$ and $v_3>v_2$.
The integrability of this situation (which is the physically relevant one on most occasions) with the corresponding choices of a Lax pair and constant parameters is illustrated in the Appendix.
However, in a situation where the group velocity $v_3$ is between the other two group velocities the system \eqref{3w} is also integrable (see the Appendix for the details).

\subsection{Conclusions and discussion}

Both the capillary-gravity waves and the currents in nature are generated by wind so that they coexist and interact in a complex way.
We have explained one possible phenomenon related to this nonlinear interaction. We have demonstrated that the $3$-wave resonance for capillary-gravity waves could take place if an underlying flow in the form of a shear is present.
Moreover, using the ``nearly'' Hamiltonian formulation in the case of constant vorticity we have extended the integrable three-wave model for this situation. The integrability is a remarkable phenomenon from both the mathematical
and the physical perspectives. From the mathematical point of view it is related to the possibility of ``solving'' the model or at least finding the very special soliton solutions (which are well known for the three waves).
The soliton solutions are believed to be very stable and preserving their shapes, velocities and energies with time. Each soliton is related to a discrete eigenvalue of the associated spectral problem.
In the case of a resonance the solitary waves are stable until/after the interaction when merger or decay takes place.  In fact, each of the functions $u_n$ in the case of a multi-soliton solution has several ``humps''
(equal to the number of the discrete eigenvalues) so that the merger/decay happens at several space-time ``points''. Illustrative figures for the one- and two-soliton solutions of the three-wave system could be found for example in \cite{GIK}.

\subsection*{Acknowledgements} C. I. Martin acknowledges the support of the Austrian Science Fund (FWF) under research grant P 30878-N32.
R. I. Ivanov gratefully acknowledges partial financial support for this work from FWF research grant P 30878-N32.
The authors are thankful for valuable comments and suggestions from three anonymous referees.

\section{Appendix}

\subsection{The operator $|D|$ and the Hilbert transform}

In order to explain the meaning of $|D|$, we introduce the Fourier transform $$\hat{u}(k):=\mathcal{F}\{u(x)\}(k),\qquad u(x)=\mathcal{F}^{-1}\{\hat{u}(k)\}(x).$$

Then $$|D|u(x):=\mathcal{F}^{-1}\{|k|\hat{u}(k)\}(x)$$ and similarly

$$\mathrm{sgn}(D)u(x):=\mathcal{F}^{-1}\{\mathrm{sgn}(k)\hat{u}(k)\}(x).$$

There is a relation between the Hilbert transform, $\mathcal{H}$

\begin{equation} \label{HT}
 \mathcal{H}\{u\} (x) := \mathrm{P.V.}\frac{1}{\pi}\int_{-\infty}^{\infty}\frac{u(x')dx'}{x-x'}
 \end{equation}

and the Fourier transforms, namely

\begin{equation}  \mathcal{F}\{\mathcal{H}\{u\}(x)\}(k) =-i \mathrm{sgn} (k) \hat{u} (k)
\end{equation} or
\begin{equation} \label{FT}\mathcal{H}\{u\}(x)=-i \mathcal{F}^{-1}\{\mathrm{sgn} (k) \hat{u} (k)\}(x).\end{equation}
Hence $$\mathcal{H}\{Du\}(x)=-i \mathcal{F}^{-1}\{|k|\hat{u} (k)\}(x)=-i |D|u(x),$$  or $$|D|=i\mathcal{H}D=\mathcal{H}\partial_x. $$

Another important property that can be deduced from \eqref{FT} is

$$\mathcal{H} e^{ikx} = -i \text{sign}(k) e^{ikx} $$

\subsection{Integration of ``slow terms'' against ``fast terms''}\label{slowafast}
We provide in this section a justification of the calculations in Section \ref{waveampl_eqs}. The investigation pertains to
integrals of the type
\begin{equation}
\mathcal{F}(a)=\int_{-\infty}^{\infty}a(X,T)e^{ikx}\,dx, \qquad k\ne 0,
\end{equation}
where $X:=\varepsilon x \simeq \mathcal{O} (\varepsilon)$ and $T:=\varepsilon t \simeq \mathcal{O} (\varepsilon)$
represent the ``slow''  variables and
$a=a(X,T)$ stands for a ``slow'' varying wave envelope integrated against a ``fast'' oscillating carrier wave with a fixed wavelength, $e^{ikx}$.
Moreover, we will also assume that
\begin{equation}\label{decayinf}
\lim_{X\to\pm\infty} a^{(n)}(X,T)=0,
\end{equation}
for all $n\in \mathbb{N}$ (here, the notation ``$^{(n)}$'' stands for the $n^{\rm th}$ derivative of $a$ with respect to $x$) and $ a^{(n)}(X,T) \simeq \mathcal{O}(1)$ for all $n\in \mathbb{N}$.

We now compute
\begin{equation}
\begin{split}
\mathcal{F}(a)=&\int_{-\infty}^{\infty}a(X,T)\p_x\left(\f{e^{ikx}}{ik}\right)\,dx=\f{a(X,T)e^{ikx}}{ik}\Big|^{X=\infty}_{X=-\infty}-\f{1}{ik}\int_{-\infty}^{\infty}e^{ikx}\p_x a(X,T)\, dx\\
=&-\f{\e}{ik}\int_{-\infty}^{\infty}e^{ikx} a^{\prime}(X,T)\,dx=\ldots=\left(\f{i\e}{k}\right)^n\int_{-\infty}^{\infty}e^{ikx}a^{(n)}(X,T)\,dx,
\end{split}
\end{equation}
for all $n\in \mathbb{N}$. The previous computation shows that $\mathcal{F}(a)$ is $\mathcal{O}(\e^n)$ for any
$n\in \mathbb{N}$. This is possible if and only if $\mathcal{F}(a)=0$. This result plays an important role in Section \ref{waveampl_eqs}.

\subsection{The three-wave equation and its integrability}

For consistency we provide some facts about the integrability of the 3-wave equation, following \cite{ZMNP}. The system arises as a compatibility between the following linear spectral problems (Lax pair):

\begin{equation}\label{LP}
\begin{split}
&-i \Psi_x=(U(x,t) + \zeta J) \Psi  \\
&-i \Psi_t=(V(x,t) + \zeta K) \Psi
\end{split}
\end{equation}
where $\zeta$ is a spectral parameter, $J=\text{diag}(J_1,J_2,J_3)$ and $K=\text{diag}(K_1,K_2,K_3)$ are constant diagonal matrices with $J_2>J_3>J_1$ and $K_1>K_3>K_2$ real constant parameters; $U:=[J,Q(x,t)]$, $V:=[K,Q(x,t)]$, where

$$Q = \begin{pmatrix}
 0 & \frac{iu_1}{\sqrt{J_2-J_1}} &  \frac{iu_3}{\sqrt{J_3-J_1}} \\
  \frac{-i\ov{u}_1}{\sqrt{J_2-J_1}} & 0 &   \frac{u_2}{\sqrt{J_2-J_3}} \\
  \frac{-i\ov{u}_3}{\sqrt{J_3-J_1}} &  \frac{-\ov{u}_2}{\sqrt{J_2-J_3}} & 0 \end{pmatrix}
 .$$ The equation arising from \eqref{LP} and the compatibility condition $\Psi_{xt}\equiv \Psi_{tx}$ is

\begin{equation}\label{3wm} [J,Q_t] -[K,Q_x] +i\left[[J,Q],[K,Q]\right]=0. \end{equation}

Componentwise, introducing $$V_1=-\frac{K_1-K_2}{J_1 -J_2}>0 , \qquad V_2=-\frac{K_1-K_3}{J_1 -J_3}>0, \qquad V_3=-\frac{K_2-K_3}{J_2 -J_3}>0 $$
the equation \eqref{3wm} leads to the system

\begin{equation}\label{3w}
\begin{split}
\dot{u}_1+V_1 u_1^{\prime}=i B \ov{u}_2 u_3\\
\dot{u}_2+V_2 u_2^{\prime}=i B \ov{u}_1 u_3\\
\dot{u}_3+V_3 u_3^{\prime}=i B  u_1 u_2
\end{split}
\end{equation}

where $$B:=\frac{(K_2-K_3)J_1+(K_3-K_1)J_2+(K_1- K_2)J_3 }{\sqrt{(J_2-J_1)(J_3-J_1)(J_2-J_3)}}  $$ is the coupling constant of the interaction. This constant could be scaled out by the obvious transformation $u_k \rightarrow u_k/B $, giving

\begin{equation}\label{3wso}
\begin{split}
\dot{u}_1+V_1 u_1^{\prime}=i  \ov{u}_2 u_3\\
\dot{u}_2+V_2 u_2^{\prime}=i  \ov{u}_1 u_3\\
\dot{u}_3+V_3 u_3^{\prime}=i  u_1 u_2
\end{split}
\end{equation}

If we take, for example, $0> K_1>K_3>K_2,$ $J_n>0$ and $K_n=-J_n^2,$ ($n=1,2,3$), then $V_3=J_2+J_3,$  $V_2=J_1+J_3$ and $V_1=J_1+J_2$ thus $V_3>V_1$ and $V_3>V_2$.

Let us mention finally that a different choice, namely $J_1>J_2>J_3$ and

$$Q = -\begin{pmatrix}
 0 & \frac{u_1}{\sqrt{J_1-J_2}} &  \frac{u_3}{\sqrt{J_1-J_3}} \\
  \frac{-\ov{u}_1}{\sqrt{J_1-J_2}} & 0 &   \frac{u_2}{\sqrt{J_2-J_3}} \\
  \frac{-\ov{u}_3}{\sqrt{J_1-J_3}} &  \frac{-\ov{u}_2}{\sqrt{J_2-J_3}} & 0 \end{pmatrix}$$
leads to exact same system \eqref{3wso}. However, noting that
\begin{equation}
\frac{V_2-V_3}{V_1-V_3}=\frac{-V_{23}+V_{13}}{-V_{12}+V_{13}}=-\frac{J_1 - J_2}{J_2- J_3} < 0  \end{equation}
in this case the group velocity $V_3$ is between the other two group velocities \cite{ZMNP,ZM}.




\end{document}